

\input amstex
\documentstyle{amsppt}

\loadbold
\loadeurm

\NoBlackBoxes
\TagsOnRight
\CenteredTagsOnSplits

\magnification=1200

\font\headerfont=cmcsc8

\def\half{\hbox{$\textstyle{1\over 2}$}}

\def\supp{\operatorname{supp}}
\def\dist{\operatorname{dist}}

\def\odd{{\text{odd}}}

\def\even{{\text{even}}}
\def\eff{{\text{eff}}}
\def\bi{{\boldkey i}}
\def\bj{{\boldkey j}}

\rightline{IAS, November 94}
\rightline{CPT-94/P.3105}
\bigskip
\bigskip

\topmatter

\title
The staggered charge-order phase\\
of the
extended Hubbard model\\
in the atomic limit
\endtitle

\leftheadtext\nofrills
   {\headerfont C. Borgs, J. J\c edrzejewski,  R.Koteck\'y}
\rightheadtext\nofrills
   {\headerfont Staggered charge-order phase
    of the extended Hubbard model}

\author
C\. Borgs\footnotemark"${}^{1,\dag}$",
J\. J\c edrzejewski\footnotemark"${}^{2}$"
and
R\. Koteck\'y\footnotemark"${}^{3,\ddag}$"
\endauthor

\footnotetext"${}^\dag$"{DFG Heisenberg Fellow}
\footnotetext"${}^\ddag$"{On leave from Center for Theoretical  Study,
Charles University, Prague;
        partly supported by
the grants GA\v CR 202/93/0449 and GAUK 376}

\affil
{\eightit
${}^1$School of Mathematics,
Institute for Advanced Study, Princeton\\
${}^2$Institute of Theoretical Physics,
Wroc\l aw University \\
${}^3$Centre de Physique Th\'eorique, CNRS, Marseille
}
\endaffil

\address Christian Borgs \hfill\newline
Institute for Advanced Study,\hfill\newline
School of Mathematics,
Princeton, NJ 08540
\endaddress

\email borgs\@math.ias.edu \endemail

\address
Janusz J\c edrzejewski\hfill\newline
Institute of Theoretical Physics, University of Wroc\l aw
\hfill\newline
Pl\. Maksa Borna 9, 50--204 Wroc\l aw, Poland
\endaddress

\email  jjed\@ift.uni.wroc.pl \endemail

\address
Roman Koteck\'y
\hfill\newline
Center for Theoretical Study, Charles
University,\hfill\newline
 Jilsk\'a 1, 110 00 Praha 1,
Czech Republic
\hfill\newline
\phantom{18.}and
\hfill\newline
Department of
Theoretical Physics, Charles University,\hfill\newline
 V~Hole\v sovi\v
ck\'ach~2, 180~00~Praha~8, Czech Republic
\endaddress

\email kotecky\@aci.cvut.cz \endemail

\abstract

We study the phase diagram of the extended
Hubbard model in the atomic limit.
At zero temperature, the phase diagram
decomposes into six regions: three
with homogeneous phases (characterized by particle
densities $\rho=0$, $1$, and $2$ and staggered
charge density $\Delta=0$) and three
with staggered phases (characterized by the
densities  $\rho=\frac12$, $1$, and
$\frac32$ and staggered densities
$|\Delta|=\frac12$,
$1$,  and $\frac12$). Here we use
Pirogov-Sinai theory to
analyze the details of the  phase diagram of
this model at low temperatures. In particular,
 we show that for any sufficiently
low nonzero temperature the three staggered
regions
merge into one staggered region $S$,
without any phase
transitions (analytic free energy
and staggered order parameter $\Delta$) within $S$.

\smallskip
{\eightpoint\noindent{\smc Keywords.}

Hubbard model, Atomic Limit, Staggered
Charge-Order, Pirogov-Sinai Theory.}

\endabstract

\endtopmatter

\document

\vfill\eject
\head{1. Introduction}
\endhead

The theory of strongly correlated electron systems
is nowadays a subject of vigorous
research.
The interest in these systems is stimulated, to a large extent,
by attempts at explaining the mechanism of high-temperature
superconductivity \cite{MRR90, Dag94}, the phenomenon of electron
localization
in narrow-band systems \cite{IILM75} and properties of quasi
one-dimensional conductors \cite{Hub79}, to name a few.
Among the models that are most frequently studied
is the Hubbard model augmented by a
nearest neighbour interaction. This model, known
as the extended Hubbard model, is defined
by the following Hamiltonian
$$
H_t= -t \!\!\sum_{ \langle \bi,\, \bj \rangle, \sigma }
\!\!\bigl( c^{*}_{\bi,\,\sigma} c_{\bj,\,\sigma} + h.c. \bigr) +
U \sum_{\bi \in \Lambda} n_{\bi,{\sssize\uparrow}}n_{\bi,
{\sssize\downarrow}} +
W \sum_{ \langle \bi,\, \bj \rangle } n_{\bi} n_{\bj} -
\bigl( \mu + zW + \frac{U}{2} \bigr)
\sum_{\bi \in \Lambda} n_{\bi}.
\tag 1.1
$$
In (1.1) we used the following notation:
at each site ${\bi}$ of a $d$-dimensional bipartite
lattice $\Lambda$, with $z$ nearest neighbours,
there are creation and annihilation operators
$c^{*}_{\bi,\,\sigma}$ and $ c_{\bi,\,\sigma}$
of the electron with
up and down spin, $\sigma=\,\uparrow,\,\downarrow$,
 that
satisfy the canonical anticommutation relations,
while
$n_{\bi,\, \sigma}:\,= c^{*}_{\bi,\,\sigma} c_{\bi,\,\sigma}$
and
$
n_{\bi}:\,
=n_{\bi,\,{\sssize\uparrow}}
+ n_{\bi,\,{\sssize\downarrow}}
$.
The first term of the  Hamiltonian (1.1)
stands for the isotropic nearest neighbour hopping of
electrons, the second one is the familiar
on-site Hubbard interaction, the third term represents
the isotropic nearest neighbour interaction,
 and the last one the contribution of the particle
reservoir characterized by the
 chemical potential $\mu$.
We have introduced the shift $zW +\frac{U}2$ in order
to move the hole-particle symmetry
point (the half-filled band) to the value $\mu=0$.
Originally, the second and the third terms
were supposed to simulate the
effect of the Coulomb repulsion
between the electrons, hence only
positive $U$ and $W$ were considered.
Later  on, in various applications
of the model, the parameters $t$, $U$
and $W$ represented  the effective
interaction constants that take  into
account also other interactions
(for instance with phonons).
Therefore $U$ and $W$ could take
negative values as well. In this
paper $U$ will be allowed to change its
sign while $W$  always stays positive.

The so called narrow band case of
the extended Hubbard model, i.e.
$|t| \ll |U|$, is of  special interest
in physical applications of the
model. It has been studied by means
of various approximate methods in
many papers (see for instance
\cite{Lor82} and references quoted
there). These studies revealed
the existence of staggered charge-order
at the hole-particle symmetry point.
The staggered charge-order is
characterized by a nonvanishing
order parameter
$$
\Delta = \lim_{\Lambda \nearrow \infty}
 |\Lambda|^{-1} \sum _{\bi \in \Lambda}
\varepsilon_{\bi} \langle n_{\bi} \rangle,
\tag 1.2
$$
where $\varepsilon_{\bi}$ assumes two values, $1$ or $-1$, depending on
to which sublattice of the bipartite lattice $\Lambda$ the site $\bi$
belongs, and $\langle \cdot \rangle$ stands for the Gibbs state.

Rigorously, the existence of staggered charge order has so far
only been established in the so called atomic limit $t\to 0$
\cite{J\c ed94}. While the above mentioned approximate results suggest
that the staggered charge order persists in the corresponding
narrow band model, the methods used in \cite{J\c ed94}
unfortunately do not allow to establish this rigorously,
because they rely on the reflection positivity of the atomic limit
model which fails to be true  for nonzero $t$.

Here we propose to study the atomic limit of the model (1.1)
using a different strategy, based on the by now classical
methods of Pirogov and Sinai \cite{PS75}, see also
\cite{Zah84,BI89}. On the one hand, these methods will
allow us to study detailed properties of the low temperature
phase diagram in the atomic limit, on the other hand they
allow for an extension to nonzero $t$, treating the narrow
band Hubbard model as a quantum perturbation of the
$t=0$ model, and leading to the rigorous proof
of staggered charge order in the narrow band Hubbard model
\cite{BK94}.

In the atomic limit, it is
convenient to rewrite the Hamiltonian (1.1) in a form
that makes the hole-particle symmetry
apparent. Namely, introducing
$Q_{\bi}:\,= n_{\bi} - 1$, we have
$$
H= \lim_{t\to 0} H_t=  \sum_{  \langle \bi,\, \bj \rangle } Q_{\bi}
Q_{\bj} +
\frac{U}{2} \sum_{\bi \in \Lambda} Q^{2}_{\bi}
 - \mu \sum_{\bi \in \Lambda} Q_{\bi},
\tag 1.3
$$
where we passed to dimensionless parameters
$U$ and $\mu$, setting $W=1$. Note that in
the atomic limit all operators appearing in the
Hamiltonian commute. Therefore,
the model (1.3) can be viewed as a two-component classical lattice
gas or, equivalently, as the classical gas  with four possible states
$0,\,\uparrow,\, \downarrow,2$ in each site,
that correspond to an empty site, a singly occupied site with spin
$\uparrow$ or $\downarrow$, and a doubly occupied
site, respectively.

\midinsert
\vskip 100truemm
\noindent
{\hbox{\includegraphics{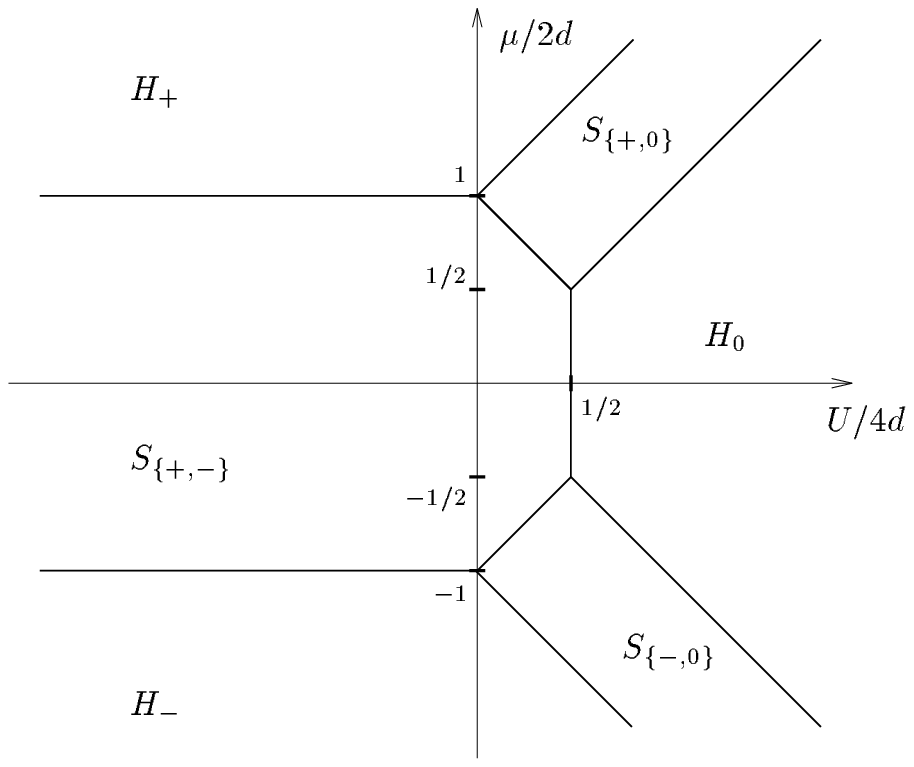}}}
\botcaption {Fig\.1. Ground state phase diagram} There are three
open regions $H_-$, $H_0$ and  $H_+$
with no staggered charge order ($\Delta=0$),
and three open regions
$S_{\{+,0\}}$, $S_{\{+,-\}}$, and $S_{\{-,0\}}$,
with staggered charge order
$|\Delta|=1/2$, $|\Delta|=1$, and $|\Delta|=1/2$, respectively.
\endcaption
\endinsert%

In the sequel, we shall present and discuss the phase diagram of the
model (1.3) on the lattice $\Bbb Z^d$.
 The ground state phase diagram  is shown in Fig\. 1.
The $(U, \mu)$ plane decomposes
into six open regions. In each of  the regions
$H_-$, $H_0$, and $H_+$, there is a
 unique homogeneous ground
state, whose particle density $\rho$, given by
$$
\rho= \lim_{\Lambda \nearrow \Bbb Z^{d} } |\Lambda|^{-1} \sum_{\bi \in
\Lambda}
\langle Q_{\bi}+1 \rangle,
\tag 1.4
$$
equals $0$, $1$, and $2$, respectively.
The remaining three regions
$S_{\{+,-\}}$, $S_{\{+,0\}}$  and $S_{\{-,0\}}$ are characterized by
staggered  charge-order. Namely, in the region $S_{\{a,b\}}$
there are two staggered ground states, denoted $[a,b]$ and $[b,a]$,
 with  $Q_{\bi}=a$ on
one sublattice and
$Q_{\bi}=b$ on the complementary  sublattice.
Note that the staggered order parameter $\Delta$ is nonvanishing in the
whole staggered region $S$ and jumps from $|\Delta|=1$ to
$|\Delta|=\frac12$ at the boundary between
$S_{\{+,-\}}$ and $S_{\{-,0\}}$  or $S_{\{+,0\}}$.

Using reflection positivity it has been shown
\cite{J\c ed94} that the
staggered long range order in the region $S$ persists at small
temperatures
$T\equiv\frac1{k \beta}>0$,
provided one stays sufficiently far away  from  the
boundary between $S$ and the homogeneous regions $H_a$, $a=0,\pm$.
However, this does not exclude  existence of a phase transition inside
the staggered region $S$. Namely, in view of the discontinuity of
$\Delta$ at the zero-temperature transition lines,
one could expect that
$\Delta$ reveals a phase transition  at nonvanishing temperatures as
well.
Indeed, mean field arguments \cite{MRC84}
predict a first order transition
surface emerging from the zero temperature transition line
separating $S_{\{+,-\}}$ from $S_{\{+,0\}}$ and
similarly for the line separating $S_{\{+,-\}}$ from
$S_{\{-,0\}}$,

Using a suitable notion of restricted ensembles we are able to
analyze this question rigorously.
Our main result here is to show the
absence of any such phase transition, in contrast to the
above mentioned mean field results.

\proclaim{Theorem A}
Consider the complement $S$ of the union of closed homogeneous regions
$\bar H_a$, $a=0,\pm$,
$$
S=\Bbb R^d \setminus (\bar H_-\cup \bar H_0\cup \bar H_+),
$$
 and let
$$
S^{(\epsilon)}=\{x\in S\mid \dist(x,S^c)>\epsilon\}.
$$
 For $d\ge 2$ and any
$\epsilon>0$ there exists a constant
$\beta_0<\infty$ (depending on $\epsilon$ and $d$) such that, for all
$\beta_0<\beta<\infty$ and
$(U,\mu)\in S^{(\epsilon)}$,
 there are exactly two phases
\footnote{As usual, a phase is defined as an extremal Gibbs state
which is periodic in all $d$ lattice directions.},
$\langle-\rangle_{\text{\rm even}}$ and
$\langle-\rangle_{\text{\rm odd}}$.
 Moreover,   $\Delta>0$  for the phase
$\langle-\rangle_{\text{\rm even}}$, $\Delta<0$ for
$\langle-\rangle_{\text{\rm odd}}$, and
both
the free energy density, $f(\beta,U,\mu)$,
and the staggered order parameters of the two phases,
$\Delta_\even(\beta,U,\mu)$ and $\Delta_\odd(\beta,U,\mu)$,
are real
analytic functions of $U$ and $\mu$ in
$S^{(\epsilon)}$.
\endproclaim

\remark{Remark}
As mentioned before, the zero temperature
staggered order parameter $\Delta$ jumps from
$\Delta=\pm 1$ to $\Delta=\pm\half$ at the
boundary between
$S_{\{+,-\}}$ and $S_{\{+,0\}}$ or $S_{\{-,0\}}$.
It is interesting to relate this jump to the smooth
behaviour for
$T=\frac1{k \beta}>0$. As we will see in Section 3,
the crossover between these two behaviours is described
by a hyperbolic tangent. Taking, e\.g\.,
the order parameter of the even phase in
the vicinity of the boundary
between, say, $S_{\{+,-\}}$ and $S_{\{+,0\}}$,
one obtains that
$$
\Delta\sim
\frac 34+\frac 14\tanh
\Bigl(\frac{2d-\mu-U/2}{kT}\Bigr)
$$
as $T\to 0$.
\endremark

Turning to
the homogeneous regions $H_a$, $a=0,\pm$, we use  standard
Pirogov-Sinai theory to discuss the low temperature behaviour inside
the corresponding regions $H_a^{(\epsilon)}$. This enables us to prove
analyticity, unicity and translation invariance of the homogeneous
phases.

As for
the boundaries between the staggered region $S$ and the
homogeneous regions $H_a$, we note that the
zero temperature coexistence
line between $H_0$ and $S_{\{+,-\}}$ gives rise to a coexistence line
surface of the two staggered phases with the homogeneous one. With
decreasing $\beta$, this surface bends towards negative $U$, i.e., above
the ground state coexistence line between $H_0$ and $S_{\{+,-\}}$ the
corresponding homogeneous phase is stable
\cite{J\c ed94}.

The remaining part of the zero temperature boundary between $S$ and the
homogeneous regions is of similar type as the boundary between
staggered and homogeneous phases in
the antiferromagnetic Ising model. For example, at the boundary between
$S_{\{+,-\}}$ and $H_+$, it is possible to join the staggered phase
$[+,-]$, without any energy cost, to the second staggered phase $[-,+]$,
going through the homogeneous phase stable in $H_+$
(see \cite{J\c ed94} and also Section 2 below for a more detailed
explanation).
 We therefore expect that this phase boundary turns, for nonzero
temperatures, into a second order transition line. Actually, in the limit
$U\to -\infty$ the extended Hubbard model in the atomic limit becomes
equivalent to the Ising antiferromagnet (with homogeneous external field
$\mu$) where this fact was rigorously proven
\cite{KY93}. In a
similar way one expects that all other boundaries between homogeneous and
staggered phases, except for the boundary between $H_0$ and
$S_{\{+,-\}}$ already considered above, correspond to second order
transition.

We summarize our knowledge of the phase diagram of the extended Hubbard
model in the atomic limit in the following theorem (see also Fig\. 2).

\midinsert
\vskip 110truemm
\noindent
\includegraphics{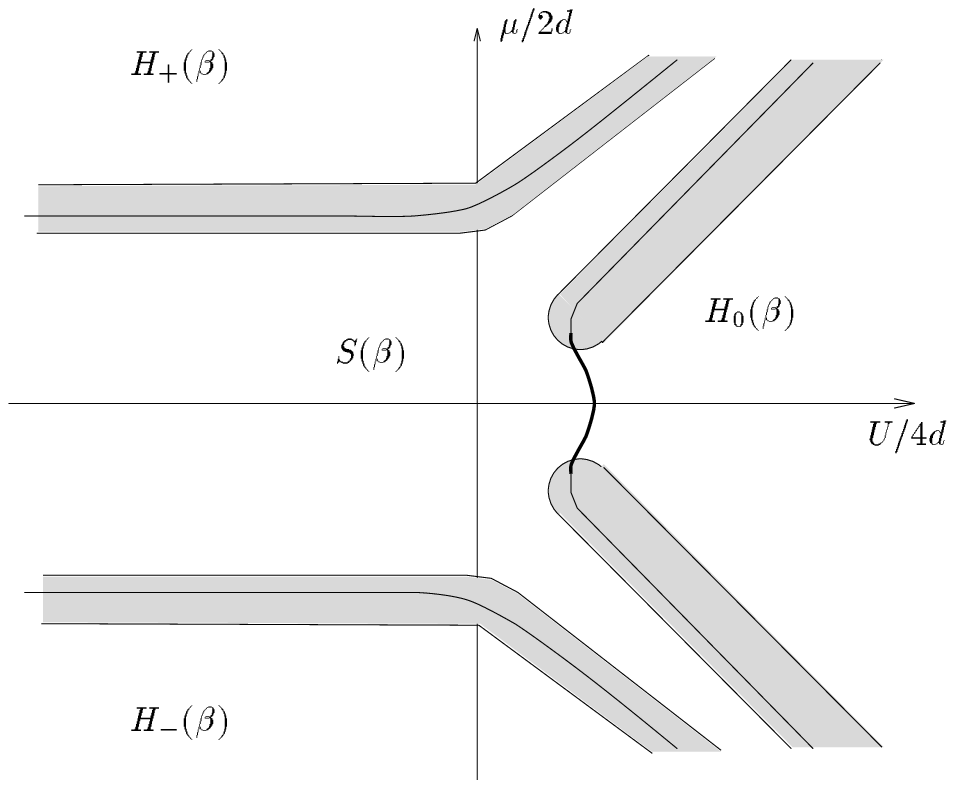}
\botcaption {Fig\.2. Phase diagram at  low temperatures}
Thin lines denote the conjectured second order transitions,
while the thick line, separating  the two phase staggered region $S$ from the
homogeneous region $H_0$,  is  first order. Shaded  are regions
over which we have no rigorous control (they shrink to the zero
temperature  lines with $\beta\to\infty$).
\endcaption
\endinsert

\bigskip

\proclaim{Theorem B}
Let $d\ge 2$ and $\beta$ be sufficiently large. Then there exist open
regions $H_a(\beta)$, $a=0,\pm$, and $S(\beta)$, where $H_0(\beta)$,
 and $S(\beta)$ touch on a curve,
$$
L(\beta)=\partial S(\beta) \cap \partial H_0(\beta)\neq\emptyset,
$$
such that the following statements are true:
\roster
\item"i)"  For $(U,\mu)\in S(\beta)$  there exist exactly two
phases,
a phase $\langle-\rangle_{\text{\rm even}}$
with  $\Delta>0$ and
a phase
$\langle-\rangle_{\text{\rm odd}}$ with $\Delta<0$.  These
phases are periodic with period 2 and
both
the free energy density, $f(\beta,U,\mu)$,
and the staggered order parameters of the two phases,
$\Delta_\even(\beta,U,\mu)$ and $\Delta_\odd(\beta,U,\mu)$,
are real analytic functions of $U$ and $\mu$ in
$S(\beta)$.
\item"ii)" For $(U,\mu)\in H_a(\beta)$, $a=0,\pm$,  there is exactly
one phase $\langle-\rangle_{a}$. For this phase, $\Delta=0$, it is
translation invariant, and  the free energy density, $f(\beta,U,\mu)$,
is a real analytic function of $U$ and $\mu$ in
in $H_a(\beta)$.
\item"iii)" On the boundary $L(\beta)$ between $S(\beta)$ and
$H_0(\beta)$, the three phases $\langle-\rangle_{a}$,
$\langle-\rangle_{\text{\rm even}}$, and
$\langle-\rangle_{\text{\rm odd}}$ coexist. Furthermore, all periodic
Gibbs states on this line are a convex combination of those three
phases.
\item"iv)" As $\beta\to\infty$, $H_a(\beta)\to H_a$, $a=0,\pm$, and
$S(\beta)\to S$.
\endroster
\endproclaim

\demo{Proof of Theorems A and B}
Theorems A and B are immediate consequences of
Propositions 1 -- 4 that are stated and proved in Section 3.
\qed
\enddemo

Before passing to the Propositions 1 -- 4 we turn to a new
representation of  the model (1.3) in terms of spin 1 variables
and then to a detailed examination of its ground state phase diagram.

\vfill
\newpage

\vfill\eject
\head{2. Structure of ground states and restricted ensembles }
\endhead

In order to rewrite the model (1.3)
in terms of a classical spin system, we
recall that all operators appearing in (1.3) commute.
However, fixing all the eigenvalues $S_{\bi}$
of the operators $Q_{\bi}$, $S_{\bi}\in\{-,0,+\}$,
does not completely specify the
state of the system, because
$S_{\bi}=0$ corresponds to  two possibilities
$n_{\bi,{\scriptscriptstyle\uparrow}}=1$
and
$n_{\bi,{\scriptscriptstyle\downarrow}}=0$
or
$n_{\bi,{\scriptscriptstyle\uparrow}}=0$
and
$n_{\bi,{\scriptscriptstyle\downarrow}}=1$.
In the partition function of the classical spin
model, this leads to a factor of $2$ for
every singly occupied site, and therefore to
an overall factor
$2^{\sum_{\bi \in \Lambda}(1-Q_{\bi}^{2})}$.
In this way the quantum system (1.3) is mapped onto
an antiferromagnetic  spin $1$ model, with
$$
H= \sum_{\langle \bi,\,\bj \rangle} S_{\bi}S_{\bj} + \frac{\tilde U}{2}
 \sum_{\bi \in \Lambda} S^{2}_{\bi} -
\mu \sum_{\bi \in \Lambda} S_{\bi},
\tag 2.1
$$
where
$$
\tilde U=U-2\beta^{-1}\ln 2.
\tag 2.2
$$
It is useful to rewrite the Hamiltonian (2.1)
 as a sum over nearest neighbour
terms
$h(S_{\bi}, S_{\bj})$, namely
$$
H=\sum_{\langle \bi,\,\bj \rangle} h(S_{\bi}, S_{\bj}),
\tag 2.3
$$
where we
introduced the energy per pair of
nearest neighbour sites
$$
h(S_{\bi}, S_{\bj})= S_{\bi}S_{\bj} + \frac{\tilde U}{4d}
( S^{2}_{\bi} +  S^{2}_{\bj} ) - \frac{\mu}{2d}
( S_{\bi} + S_{\bj}).
\tag 2.4
$$
This form of the Hamiltonian makes the
task of constructing the ground state phase diagram,
i.e. determining the six regions
$H_a$, $a=0,\pm$, $S_{ \{+,- \} }$, $S_{ \{+,0 \} }$,
 and $S_{ \{-,0 \} }$, mentioned in previous section, an easy exercise.
The energies of the nearest neighbour spin configurations are
$$
\alignat 2
&h(+,+)&&= \phantom{-}1+ {\textstyle \frac{\tilde U}{2d} -
\frac{\mu}{d}},
\tag2.5a\cr
&h(+,0)&&= \phantom{-}0+{\textstyle \frac{\tilde U}{4d} -
\frac{\mu}{2d}},
\tag2.5b\cr
&h(+,-)&&=-1+{\textstyle \frac{\tilde U}{2d}},
\tag2.5c\cr
&h(0,-)&&=\phantom{-}0+{\textstyle\frac{\tilde U}{4d} +
\frac{\mu}{2d}},
\tag2.5d\cr
&h(0,0)&&=\phantom{-}0.
\tag2.5e\cr
&h(-,-)&&= \phantom{-}1+ {\textstyle \frac{\tilde U}{2d} +
\frac{\mu}{d}},
\tag2.5f\cr
\endalignat
$$
Using (2.5), we find three homogeneous regions
$$
H_{+}:\,=
\Bigl\{
   (U,\mu)\mid  \frac{\mu}{2d} >
   \max\Bigl\{1,1+ \frac{\tilde U}{4d} \Bigr\}
\Bigr\},
\tag 2.6a
$$
with minimal energy pairs $(+,+)$,
$$
H_{-}:\,=
\Bigl\{
    (U,\mu)\mid  -\frac{\mu}{2d} >
    \max\Bigl\{1,1+ \frac{\tilde U}{4d} \Bigr\}
\Bigr\},
\tag 2.6b
$$
with  minimal energy pairs  $(-,-)$, and
$$
H_{0}:\,=
\Bigl\{
   (U,\mu)\mid
   \max\bigl\{\left|\frac{\mu}{2d}\right|,\frac12\bigr\}
 < \frac{\tilde U}{4d}
\Bigr\},
\tag 2.6c
$$
with minimal energy pairs  $(0,0)$;
and three staggered regions
$$
S_{\{ +,- \}}:\,=
\Bigl\{
    (U,\mu)\mid \left|\frac{\mu}{2d}\right|
    < \min\Bigl\{1,1 - \frac{\tilde U}{4d}\Bigr\}
    \quad\text{and}\quad
    \frac{\tilde U}{4d} < \frac{1}{2}
\Bigr\},
\tag 2.7a
$$
with minimal energy pairs $(+,-)$ and $(-,+)$,
$$
S_{\{ +,0 \}}:\,=
\Bigl\{
  (U,\mu)\mid
  \frac{\tilde U}{4d} < \frac{\mu}{2d}  < 1 + \frac{\tilde U}{4d}
  \quad\text{and}\quad
  \frac{\tilde U}{4d} < \frac{\mu}{2d}
\Bigr\},
\tag 2.7b
$$
with minimal energy pairs  $(+,0)$ and $(0,+)$, and
$$
S_{\{ -,0 \}}:\,=
\Bigl\{
  (U,\mu)\mid
  \frac{\tilde U}{4d} < -\frac{\mu}{2d}  < 1 + \frac{\tilde U}{4d}
  \quad\text{and}\quad
  \frac{\tilde U}{4d} < -\frac{\mu}{2d}
\Bigr\},
\tag 2.7c
$$
with minimal energy pairs $(- ,0)$ and $(0,- )$.
Thus, in each of the regions
$H_{a}$, $a=0,\pm$, there is
a  unique homogeneous ground configuration
$\{ S_{\bi}=a \}_{\bi \in \Lambda}$.
On the other hand, in each of the regions
$S_{\{ a,b \}}$ there are exactly two
 ground configurations, such that, when on one
sublattice $S_{\bi}=a$, on
the complementary one $S_{\bi}=b$, and vice versa.
In order to relate the formulae (2.6) and (2.7) to Fig\. 1, we
notice that $U=\tilde U$ for $\beta=\infty$.

At this moment, let us remark that the above analysis of ground
configurations shows that the components $h$ of $H$ (cf\. (2.4))
constitute a so called $m$--potential
\cite{Sla87}.
Moreover, since in each of the regions $H_{+}$, $H_{0}$, $H_{-}$,
$S_{\{ +,-\}}$,
$S_{\{ +,0 \}}$, and $S_{\{- ,0 \}}$ there are only finitely many
ground  configurations, one can readily  apply
standard Pirogov--Sinai theory to study the
low temperature properties of the
corresponding phases, away from the boundaries of  these regions.

On the boundaries of the regions $H_{+}$, $H_{0}$, $H_{-}$,
$S_{\{+,-\}}$,
$S_{\{ +,0 \}}$, and $S_{\{ -,0 \}}$, the structure of the ground
states  is  more complicated.
Combining the minimal energy pair configurations of the
corresponding  adjacent regions, one can construct infinitely many
 ground configurations everywhere, except for the
boundary between
$S_{\{ +,-
\}}$ and $H_{0}$. On the latter boundary
there are exactly three ground configurations, namely, those
that correspond to the adjacent regions. This, of course, places
also this case into the realm of standard Pirogov-Sinai theory.

There is an important difference
between the infinitely degenerated  boundaries
shared by staggered and homogeneous regions
and those shared  by two staggered
regions. In the first case, i.e. on the boundary between, say, $H_{+}$
and $S_{\{+,-\}}$, the minimal energy pairs of both regions, namely
the pairs
$(+,+)$, $(+,-)$ and $(-,+)$, can be combined into an arbitrary
configuration  made out of ``$+$''
and ``$-$'', as long as no nearest neighbour pair of minuses is present.
Mutatis mutandis, the same is true for the other infinitely degenerate
boundaries between
staggered and homogeneous regions.
One therefore
obtains the same structure of ground states as in the Ising antiferromagnet
at the critical field, presumably giving rise
to surfaces of second order transitions at nonzero temperature.

In the
second case, i.e. on the boundaries shared by two staggered
regions,  the situation is different.
Considering  for instance the region $S_{+}$ that consists of
$S_{\{ +,-\}}$,  $S_{\{ +,0 \}}$, and the
boundary shared by these regions,
we introduce two disjoint classes of configurations,
$\Cal G^+_{\text{even}}$ and $\Cal G^+_{\text{odd}}$.
Namely, we define $\Cal G^+_{\text{even}}$  as the set
 of all configurations, for which $S_{\bi}=+$ on the even
sublattice,  while
$S_{\bi}=0$ or $-$ on the odd sublattice, and the set
$\Cal G^+_{\text{odd}}$  by interchanging the role of the two
sublattices.  Then, for all points in $S_+$, every ground
configuration falls into one of these two disjoint classes. This
already suggests to use, in $S_+$, a version of Pirogov-Sinai theory
with the sets $\Cal G^+_{\text{even}}$ and $\Cal G^+_{\text{odd}}$
playing the role of restricted ensembles
\cite{BKL85} that replace the ground
states\footnote{Notice that different
configurations from $\Cal G^+_{\text{even}}$
(and similarly $\Cal G^+_{\text{odd}}$)
attain, in general, different
energies.
In particular, only on the boundary between
$S_{\{ +,-\}}$ and $S_{\{ +,0 \}}$,
all configurations from $\Cal G^+_{\text{even}}$ are ground
configurations. As a consequence, the version of the Pirogov-Sinai theory for
lattice systems with residual entropy \cite{GS88, SGL89} does not apply here.
Namely, it would need in our case an assumption that everywhere in the
coexistence region $S_+$ all configurations from $\Cal G^+_{\text{even}}$ (and
$\Cal G^+_{\text{odd}}$) are ground configurations.}.
The same
remark applies to the region $S_-$ made of
$S_{\{ +,-\}}$,  $S_{\{ -,0 \}}$ and their shared boundary,
and the corresponding sets
$\Cal G^-_{\text{even}}$ and $\Cal G^-_{\text{odd}}$.

\vfill\eject
\head{3. Proof of Theorems A and B}
\endhead

As mentioned above, Theorems A and B
are an immediate consequence of the following four
Propositions.

\proclaim{Proposition 1}
Consider the regions
$$
H_a^{(\epsilon)}=\{x\in H_a\mid \dist(x,H_a^c)>\epsilon\},
\qquad a=0,\pm .
$$
For $d\ge 2$ there exists a constant
$b=b(d)>0$, such that
for all $\epsilon < \infty $,
$\beta>b/\epsilon$, and
 $(U,\mu)\in H_a^{(\epsilon)}$, there is
exactly one
phase $\langle-\rangle_{a}$. For this phase
$\Delta=0$, it is translation invariant, and  the free energy
density $f(\beta,U,\mu)$  is
analytic  in
$$
{\Cal H}_a^{(\epsilon)}=
\biggl\{
(\beta,U,\mu)\in\Bbb C^3
\mid
\operatorname{Re}\,\beta\geq b/\epsilon,
\biggl(\frac{\operatorname{Re}\,\beta U}{\operatorname{Re}\,\beta},
       \frac{\operatorname{Re}\,\beta\mu}{\operatorname{Re}\,\beta}
\biggr)
\in H_a^{(\epsilon)}
\biggr\}.
\tag 3.1
$$
\endproclaim

\demo{Proof}
Except for the last statement, the proposition follows
immediately from standard Pirogov-Sinai theory
\cite{PS75, Mar75, Zah84}:
Given the relations (2.5) and (2.6), one gets,
 for a suitable constant
$\alpha=\alpha(d)>0$
and
$(U,\mu)\in H_a^{(\epsilon)}$,
the inequality
$$
\beta h(b,c)
\geq
\beta h(a,a)
+\beta\alpha\epsilon
\tag 3.2
$$
provided
$(b,c)\neq (a,a)$.
As a consequence, all excitations of the ground state
$(a,a)$ are exponentially suppressed,
with a ``Peierls constant'' $\tau=\tilde\alpha \beta\epsilon$
where $\tilde\alpha>0$
depends only on the dimension $d$.

In order to prove the last statement of the proposition,
one needs a representation in terms of a convergent
contour expansion inside
$\Cal H_a^{(\epsilon)}$,
where the Hamiltonian $H$ is complex.
This situation has been dealt with in \cite{BI89}. In order
to prove the corresponding Peierls condition,
one needs a relation of the form (3.2) for the
{\it real part}
of $\beta H$, namely
$$
\operatorname{Re}\,
\beta h(b,c)
\geq
\operatorname{Re}\,\beta h(a,a)
+(\operatorname{Re}\,\beta)\alpha\epsilon
\,.
\tag 3.3
$$
This leads to the regions $\Cal H_a^{(\epsilon)}$.
\qed
\enddemo

\proclaim{Proposition 2}
Consider the regions $S_{\{a,b\}}$ and the corresponding sets
$
S_{\{a,b\}}^{(\epsilon)}
$,
${\{a,b\}}={\{+,-\}},\;\{+,0\},\;\{-,0\}$.
For $d\ge 2$ there exists a constant
$b=b(d)>0$, such that
for all $\epsilon < \infty $,
$\beta>b/\epsilon$, and
 $(U,\mu)\in S_{\{a,b\}}^{(\epsilon)}$, there exist exactly two
phases, the  phase
$\langle-\rangle_{\text{\rm even}}$
with  $\Delta>0$ and the
phase
$\langle-\rangle_{\text{\rm odd}}$
with $\Delta<0$.
These
phases are periodic with period 2 and  the free energy density,
$f(\beta,U,\mu)$,
as well as the corresponding staggered order parameters,
$\Delta_\even(\beta,U,\mu)$ and $\Delta_\odd(\beta,U,\mu)$,
are analytic  in
$$
\Cal S_{\{a,b\}}^{(\epsilon)}=
\biggl\{
(\beta,U,\mu)\in\Bbb C^3
\mid
\operatorname{Re}\,\beta\geq b/\epsilon,
\biggl(\frac{\operatorname{Re}\,\beta U}{\operatorname{Re}\,\beta},
       \frac{\operatorname{Re}\,\beta\mu}{\operatorname{Re}\,\beta}
\biggr)
\in S_{\{a,b\}}^{(\epsilon)}
\biggr\}.
\tag 3.4
$$
\endproclaim

\demo{Proof}
Again, the proof is standard, except for the last statement.
Actually, in view of the essential singularities
associated with first-order phase transitions
\cite{Isa84},
the analyticity proof
in the coexistence regions
$\Cal S_{\{a,b\}}^{(\epsilon)}$
is more subtle than that in the
single phase regions
$\Cal H_a^{(\epsilon)}$.

We start with the observation that a Peierls condition
of the form (3.3), namely
$$
\operatorname{Re}\,
\beta h(c,d)
\geq
\operatorname{Re}\,\beta h(a,b)
+(\operatorname{Re}\,\beta)\alpha\epsilon
\qquad\text{for all}\qquad
\{c,d\}\neq\{a,b\}
\,,
\tag 3.5
$$
is valid in all of
$\Cal S_{\{a,b\}}^{(\epsilon)}$.
Introducing truncated contour models as in
\cite{Zah84, BI89} to expand about
the two staggered ground states
$[a,b]$ and $[b,a]$,
one therefore obtains a convergent cluster
expansion for the corresponding
``truncated free energies''
$f_{\text{even}}$ and $f_{\text{odd}}$.
Next we note that
$$
f_{\text{even}}(\beta,U,\mu)
=f_{\text{odd}}(\beta,U,\mu)
\qquad\text{for all}\qquad
(\beta,U,\mu)
\in\Cal S_{\{a,b\}}^{(\epsilon)}
\,
$$
due to the translation symmetry of the
model. As a consequence,
$$
\text{Re}\,
(\beta f_{\text{even}}(\beta,U,\mu))
=
\text{Re}\,
(\beta
f_{\text{odd}}(\beta,U,\mu))
\qquad\text{for all}\qquad
(\beta,U,\mu)
\in\Cal S_{\{a,b\}}^{(\epsilon)}
\,.
$$
The results of \cite{BI89}
then imply that both the even and the odd
phase are stable in all of
$\Cal S_{\{a,b\}}^{(\epsilon)}$,
implying in particular that
the truncated free energies are
equal to the corresponding ``physical free energy''
$f(\beta,U,\mu)$
obtained as the limit of (logarithms of)
finite volume partition functions.
As a consequence,
$f(\beta,U,\mu)$ can be expressed as an absolutely
convergent sum of analytic terms,
implying that $f(\beta,U,\mu)$ is analytic itself.
The stability of both the
even and the odd phase also implies the convergence
of the contour expansion for the staggered order
parameters
$\Delta_\even(\beta,U,\mu)$ and $\Delta_\odd(\beta,U,\mu)$,
yielding their analyticity in
$
\Cal S_{\{a,b\}}^{(\epsilon)}
$.
\qed
\enddemo

\remark{Remark}
Let us note the differences between the
situation of Proposition 2 and the phase coexistence of,
say, an Ising ferromagnet at zero field	 $h$. In the situtation
of Proposition 2, where the two phases $\langle - \rangle_\even$
and  $\langle - \rangle_\odd$ can be obtained from each other by
a translation, $f_\even=f_\odd$ troughout the complex
region
$
\Cal S_{\{a,b\}}^{(\epsilon)}
$,
a statement which is stronger than the
stability condition
$
\text{Re}\,(\beta f_{\text{even}})
=
\text{Re}\,(\beta f_{\text{odd}})
$.
For the Ising model, on the other hand,
the symmetry relating the two phases
$\langle - \rangle_+$    and $\langle - \rangle_-$
requires a change of the sigh of $h$. As a
consequence,
no open neighbourhood $\Cal U\subset\Bbb C$
of $h=0$ can be found such that both the plus and
the minus phase are stable in $\Cal U$.
Furthermore, on the coexistence line
$\text{Re}\,h=0$
where both phases are stable,
$f_+\not\equiv f_-$, even though the weaker condition
$
\text{Re}\,(\beta f_+)
=
\text{Re}\,(\beta f_-)
$
is true for
$\text{Re}\,h=0$.
\endremark

\proclaim{Proposition 3}
Let $R$ be the union of
$H_0$, $S_{\{+,-\}}$ and their common boundary,
and let
$
R^{(\epsilon)}=
\{x\in R\mid \dist(x,R^c)>\epsilon\}
$.
For $d\ge 2$ there exists a constant
$b=b(d)>0$, such that
for all $\epsilon < \infty $ and
$\beta>b/\epsilon$
there exist a curve $L(\beta)$,
separating $R^{(\epsilon)}$ into two open regions
$R_0^{(\epsilon)}$ and $R_{\{+,-\}}^{(\epsilon)}$,
such that the following statements are true.

\item{i)} In $R_0^{(\epsilon)}$,
there exits exactly one phase $\langle-\rangle_0$.
This phase is translation invariant with $\Delta=0$.

\item{ii)} In $R_{\{+,-\}}^{(\epsilon)}$
there are exactly two phases
$\langle-\rangle_{\text{even}}$
and
$\langle-\rangle_{\text{odd}}$,
characterized by $\Delta>0$ and $\Delta<0$.
Both phases are periodic with period $2$.

\item{iii)} On the boundary $L(\beta)$
between $R_0^{(\epsilon)}$
and $R_{\{+,-\}}^{(\epsilon)}$,
all three phases
$\langle-\rangle_0$,
$\langle-\rangle_{\text{even}}$
and
$\langle-\rangle_{\text{odd}}$
coexist. Furthermore, all periodic Gibbs
states on this curve are a convex combination of these
three phases.

\item{iv)} As a function of $U$ and $\mu$,
the free energy $f$ is real analytic in
$R^{(\epsilon)}\setminus L(\beta)$, and the
staggered order parameters of the two phases
$\langle-\rangle_{\text{even}}$
and
$\langle-\rangle_{\text{odd}}$,
$\Delta_\even(\beta,U,\mu)$ and $\Delta_\odd(\beta,U,\mu)$,
are real analytic in
$R_{\{+,-\}}^{(\epsilon)}$.
\endproclaim

\demo{Proof}
Again the proof is standard. One now introduces
three different truncated contour models:
one for the excitations about the homogeneous
configuration $(0,0)$ and two for the excitations
about the two staggered configurations $(+,-)$ and
$(-,+)$. In the region $R^{(\epsilon)}$,
and more generally in the complex region
$$
\Cal R^{(\epsilon)}=
\{(U,\mu)\mid
(\text{Re}\,U,\text{Re}\,\mu)
\in R^{(\epsilon)}\}
\,,
\tag 3.6
$$
the model again satisfies a suitable
Peierls condition provided
$\beta\epsilon$ is big enough. This leads
to the convergence of the cluster expansion for
the corresponding truncated free energies
$f_0$, $f_{\text{even}}$ and
$f_{\text{odd}}$ in
$\Cal R^{(\epsilon)}\supset R^{(\epsilon)}$.

Given the ``degeneracy removing condition''
$$
\frac d{dU}
\biggl(
h(+,-)- h(0,0)
\biggr)
=\frac 1{2d}>0
\tag 3.7
$$
and the symmetry relation
$$
f_{\text{even}}(U,\mu)=
f_{\text{odd}}(U,\mu)
\,,
\tag 3.8
$$
the proof of statement i) -- iii) is now
an easy application of the methods
of \cite{Zah84}.
Actually, the complex analogue of (3.7), namely
the degeneracy removing condition
$$
\frac d{d\text{Re}\,U}
\biggl(
\text{Re}\, h(+,-)
-\text{Re}\, h(0,0)
\biggr)
=\frac 1{2d}>0
\,,
\tag 3.9
$$
together with the validity of (3.8) in
the complex region
$\Cal R^{(\epsilon)}$
leads to the existence of a phase
transition surface
$\Cal S(\beta)$ that separates
$\Cal R^{(\epsilon)}$ into two open regions:
a region $\Cal R_0^{(\epsilon)}$
in which
$
\text{Re}\, (\beta f_0(U,\mu))
<
\text{Re}\,(\beta f_{\text{even}}(U,\mu))
$
and $f(U,\mu)=f_0(U,\mu)$,
and a region
$\Cal R_{\{+,-\}}^{(\epsilon)}$
in which
$
\text{Re}\, (\beta f_0(U,\mu))
>
\text{Re}\,(\beta f_{\text{even}}(U,\mu))
$
and
$f(U,\mu)=f_{\text{even}}(U,\mu)$,
see \cite{BI89}
\footnote{In the language of \cite{Zah84, BI89},
 $\Cal R_0^{(\epsilon)}$
is the region where the homogeneous phase is
stable,
and $\Cal R_{\{+,-\}}^{(\epsilon)}$
is the region where the two staggered
phases are stable.}.
As a consequence, the free energy $f$ of
the model can be rewritten as
a convergent sum of analytic terms
in both $\Cal R_0^{(\epsilon)}$
and
$\Cal R_{\{+,-\}}^{(\epsilon)}$,
leading to the analyticity
of $f$ in
$
\Cal R^{(\epsilon)}
\setminus
\Cal L(\beta)
$
and hence the real analyticity of $f$
in $R^{(\epsilon)}\setminus L(\beta)$.
In a similar way, one obtains the real
analyticity of the charged order parameters
$\Delta_\even(\beta,U,\mu)$ and $\Delta_\odd(\beta,U,\mu)$.
\qed
\enddemo

\proclaim{Proposition 4}
Consider the regions $S_\pm$ introduced
in the last section, and the corresponding
sets $S_\pm^{(\epsilon)}$. For $d\ge 2$
and $m=\pm$ there exists a constant
$b=b(d)>0$, such that
for all $\epsilon < \infty $,
$\beta>b/\epsilon$, and
$(U,\mu)\in S_m^{(\epsilon)}$,
there exist exactly two
phases, a phase
$\langle-\rangle_{\text{\rm even}}$
with  $\Delta>0$ and a
phase $\langle-\rangle_{\text{\rm odd}}$
with $\Delta<0$.
These
phases are periodic with period 2 and
both the free energy density,
$f(\beta,U,\mu)$, and the corresponding
staggered order parameters,
$\Delta_\even(\beta,U,\mu)$ and $\Delta_\odd(\beta,U,\mu)$,
are real analytic functions of $U$ and $\mu$
in $S_m^{(\epsilon)}$.
\endproclaim

\demo{Proof of Proposition 4}
Without loss of generality, we may assume that
$(U,\mu)\in S_+$. In order to prove
the proposition, we introduce an auxiliary
Ising variable $\sigma_\bi=\sigma(S_\bi)$
as
$$
\sigma(S_\bi)=
\cases +&\text{if}\quad S_\bi=+\\
       -&\text{if}\quad S_\bi\in\{0,-\}
\endcases
\,,
\tag 3.10
$$
and rewrite the model (2.1) in terms of an
effective Hamiltonian $H^\eff(\sigma)$.
We then  prove that the Hamiltonian
$H^\eff$ has two  ground states
$g^{\text{even}}$ and $g^{\text{odd}}$,
corresponding to the restricted ensembles
$\Cal G_{\text{even}}^+$ and $\Cal G_{\text{odd}}^+$
introduced
at the end of the last section,
$$
g_\bi^{\text{even}}=\epsilon_\bi
\quad\text{and}\quad
g_\bi^{\text{odd}}=-\epsilon_\bi
\,,
\tag 3.11
$$
and that the excitation above these
ground states obey a suitable Peierls
condition. Here, as in Section 1,
$\epsilon_\bi=+1$ on the even
and
$\epsilon_\bi=-1$ on the odd sublattice.

We start with some notation.
We consider a box
$\Lambda=[1,L]^d\cap\Bbb Z^d$,
its boundary
$
\partial\Lambda=
\{\bi\in\Lambda^c\mid \dist(\bi,\Lambda)=1\}
$,
the set $B(\Lambda)$ of nearest
neighbour bonds $\langle\bi,\bj\rangle$
with at least one endpoint in $\Lambda$,
and the union of $\Lambda$ and its
boundary,
$\bar\Lambda=\Lambda\cup\partial\Lambda$.
Here, as in the sequel, $\dist(\cdot,\cdot)$
denotes the $\ell_1$ distance in $\Bbb Z^d$.
As usual, we call two sets
$V$, $V^\prime\in\Bbb Z^d$
adjacent or touching
if $\dist(V,V^\prime)=1$,
and a set $V\subset\Bbb Z^d$ connected
if for any two points $\bi,\bj\in V$
there is a sequence of adjacent points
in $V$ that joins $\bi$ to $\bj$.

Keeping in mind that we want to
construct finite temperature
states $\langle  - \rangle_m$
which are small perturbations of the
restricted ensembles $\Cal G_m^{+}$,
$m=$ ``even'' or ``odd'', we introduce
an
effective Hamiltonian
$H_\Lambda^\eff(\sigma_\Lambda\mid m)$ in $\Lambda$ with the
boundary conditions $m=$ even, odd, by
$$
e^{-\beta H_\Lambda^\eff(\sigma_\Lambda\mid m)}
=\sum\Sb \scriptstyle S_{\bar\Lambda}:\,
\matrix\format\l\\
\scriptstyle \sigma(S_\bi)=\sigma_\bi, \bi\in\Lambda
\\
\scriptstyle \sigma(S_\bi)=g^{m}_\bi, \bi\in\partial\hskip -.03cm \Lambda
\endmatrix\endSb
\;\prod_{\langle\bi,\bj\rangle\in B(\Lambda)}
e^{-\beta h(S_\bi,S_\bj)}
\,.
\tag 3.12
$$
The
corresponding finite volume Gibbs states are
$$
\langle\, \cdot\, \rangle_{m,\Lambda}
=\frac 1{Z_m(\Lambda)}
\sum_{\sigma_\Lambda}
\,\,\cdot\,\,e^{-\beta H_\Lambda^\eff(\sigma_\Lambda\mid m)}
\tag 3.13
$$
with
$$
Z_m(\Lambda)=\sum_{\sigma_\Lambda}
e^{-\beta H_\Lambda^\eff(\sigma_\Lambda\mid m)}.
\tag 3.14
$$
Extending the configuration $\sigma_\Lambda$
to $\bar\Lambda$ by setting
$\sigma_\bi=g_\bi^{m}$ for
$\bi\in\partial\Lambda$,
we define a nearest neighbour pair
$\langle\bi,\bj\rangle
\in B(\Lambda)$
as {\it excited}
in the configuration $\sigma_\Lambda$
if $\sigma_\bi=\sigma_\bj$ and
a point $\bi\in\bar\Lambda$
as {\it excited} if it is contained in
an excited bond. Note that
the notion of whether a bond
$\langle\bi,\bj\rangle$
that joins the volume $\Lambda$ to its
boundary $\partial\Lambda$ is
excited or not depends on the
boundary condition.

At this point, contours and ground state
regions are defined in
the standard way: Given a configurations
$\sigma_\Lambda$ (and one of the two
boundary conditions introduced above),
the {\it contours} $Y_1,\cdots,Y_n$
corresponding to the
configuration $\sigma_\Lambda$ are defined
as
pairs of the form $Y=(\supp Y, \sigma_Y)$,
where $\supp Y$ is a connected component
of the set of excited points and
$\sigma_Y$ is the restriction of
$\sigma_{\bar\Lambda}$ to $\supp Y$.
The {\it ground state regions}
are defined as the connected components
of the set of points which are not exited.
Note that the restriction of $\sigma_{\bar\Lambda}$
to a ground state region $C$ is staggered,
and hence equal to the restriction of one
of the two ground states to $C$,
$\sigma_C=g^{m}_C$, where $m=m(C)$
may vary from component to component.

An important property of a set of contours
$Y_1,\cdots,Y_n$
corresponding to a
configuration $\sigma_\Lambda$
is that they ``match''.
In order to define this notion,
we note that each contour $Y$
determines the value of $\sigma_\bi$
on all points $\bi\in\bar\Lambda$
which touch its support
because all bonds joining
the support of $Y$ to such a point
are not exited. The contour $Y$ therefore
determines the value of $m(C)$ for all
ground state regions $C$ touching
its support. We say that $Y$ attaches a
{\it label} $m(C)=m_Y(C)$ to these
ground state regions.
{\it Matching} of the contours
$Y_1,\cdots,Y_n$ is the statement
that the labels attached to a given
ground state region $C$ by different
contours are identical and compatible with
the boundary condition.
A minute of reflection now shows
that to each set $\{Y_1,\cdots,Y_n\}$
of matching contours with
$\dist(\supp Y_k,\supp Y_l) > 1$,
$k\neq l$, there corresponds exactly
one configuration $\sigma_\Lambda$.
The partition function $Z_m(\Lambda)$
can therefore be expressed as a sum
over sets of matching contours,
once the Hamiltonian $H_\Lambda^\eff(\sigma_\Lambda\mid m)$
has been expressed in terms of
$Y_1,\cdots,Y_n$.

We will now show that this can be done in the form
$$
e^{-\beta H_\Lambda^\eff(\sigma_\Lambda\mid m)}
=e^{-\beta H_\Lambda^\eff(g^m_\Lambda\mid m)}
\prod_{k=1}^n z(Y_k)
\tag 3.15
$$
where $z(Y_k)$ are contour
weights obeying a Peierls condition
$$
|z(Y_k)|\leq e^{-\tau|\supp Y|}
\,
\tag 3.16
$$
with sufficiently large Peierls constant
$\tau$.

We start with an explicit calculation of the
Hamiltonian
$ H_\Lambda^\eff(\sigma_\Lambda\mid m)$
for the configuration
$\sigma_\Lambda=g^{m}_\Lambda$
with no contour.
In this configuration,
each point $\bi\in\bar\Lambda$
with $\sigma_\bi=-$
has $2d$ nearest neighbours
$\bj\in\bar\Lambda$
with $\sigma_\bj=+$
if $\bi\in\Lambda$,
and $1$ nearest neighbour
$\bj\in\bar\Lambda$
with $\sigma_\bj=+$
if $\bi\in\partial\Lambda$.
Since $\sigma_\bj=\sigma(S_\bj)=+$
implies
$S_\bj=+$,
the summation over
the spin variable $S_\bi$
in (3.12)
therefore
leads to a factor
$$
\lambda=\sum_{S_\bi:\sigma(S_\bi)=-}
e^{-2d\beta h(S_\bi,+)}
=e^{-2d\beta h(0,+)}+e^{-2d\beta h(-,+)}
\tag 3.17
$$
if $\bi\in\Lambda$, and to a factor
$$
\lambda^\prime=\sum_{S_\bi:\sigma(S_\bi)=-}
e^{-\beta h(S_\bi,+)}
=e^{-\beta h(0,+)}+e^{-\beta h(-,+)}
\tag 3.18
$$
if $\bi\in\partial\Lambda$.
Per bond, this yields the energy
$$
h_0^{m}(\langle\bi,\bj\rangle)
=\cases
   h_0^\prime=
   -\frac 1{\beta}\log\lambda^\prime
   &\text{if $\bi \in \partial\Lambda$
    and $\sigma_\bi=g_\bi^{m}=-$}
\\
   h_0=
   -\frac 1{2d\beta}\log\lambda
   &\text{if $\bi \in \Lambda$
    and $\sigma_\bi=g_\bi^{m}=-$}
\endcases
\,.
\tag 3.19
$$
For the energy of the configuration
$g^{m}_\Lambda$, this gives
$$
H_\Lambda^\eff(g^m_\Lambda\mid m)
=
{\sum_{\langle\bi,\bj\rangle\in B(\Lambda)}
h_0^{m}(\langle\bi,\bj\rangle)}
\,.
\tag 3.20
$$
\remark{Remark} Obviously, the
boundary correction (3.19) does not
affect the specific ground state energy
$$
e_m=\lim_{\Lambda\to\Bbb Z^d}
\frac {1}{|\Lambda|}
H_\Lambda^\eff(g^m_\Lambda\mid m)
\,
\tag 3.21
$$
so that $e_\even=e_\odd$. It does
affect, however, the finite volume
ground state energies
$H_\Lambda^\eff(g^m_\Lambda\mid m)$.
\endremark

In order to calculate the weight
$e^{-\beta H_\Lambda^\eff(\sigma_\Lambda\mid m)}$
for a configuration
$\sigma_\Lambda$ corresponding to
a
nonempty set of  contours
$\{Y_1\cdots,Y_n\}$,
we  extract from (3.12), for each contour
$Y\in\{Y_1\cdots,Y_n\}$, the factor
$$
\tilde z(Y)=
\sum_{S_Y:\sigma(S_i)=\sigma_i}
\;
\prod_{\langle\bi,\bj\rangle\in B(Y)}
e^{-\beta h(S_\bi,S_\bj)}
\,.
\tag 3.22
$$
 Here $B(Y)$ is the set of all
bonds $\langle\bi,\bj\rangle\in B(\Lambda)$
such that either
\item{i)} both endpoints of
 $\langle\bi,\bj\rangle$ are in the support of $Y$,

\noindent
or
\item{ii)} only one endpoint of
$\langle\bi,\bj\rangle$ lies in the support of
$Y$, and this endpoint corresponds to a value
$\sigma_\bi=-1$.

\noindent
Note that the second class of bonds are
those bonds which couple the
spin variables in the support of $Y$ to
the spin variables in
$\bar\Lambda\setminus\supp Y$.
The remaining sum in (3.12) can
be easily calculated because
all points
$\bi\in\Lambda\setminus
(\supp Y_1\cup\cdots\cup
\supp
Y_n)$ with
$\sigma_\bi=-$ are not
excited. The summation over the corresponding
spin variable $S_\bi$ therefore
again leads to factors $\lambda$
and $\lambda^\prime$, giving a factor
$e^{-\beta h_0^{m}(\langle\bi,\bj\rangle)}$
for all the bonds in
$B(\Lambda)\setminus(B(Y_1)\cup\cdots\cup B(Y_n))$.
Extracting the factor
$$
\prod_{\langle\bi,\bj\rangle\in B(Y)}
e^{-\beta h_0^{m}((\langle\bi,\bj\rangle))}
$$
{}from the activities (3.22), we therefore
obtain a representation of the form
(3.15), with
$$
z(Y)=\sum_{S_Y:\sigma(S_i)=\sigma_i}
\;\prod_{\langle\bi,\bj\rangle\in B(Y)}
e^{-\beta
(h(S_i,S_j)-h_0^{m}(\langle\bi,\bj\rangle))}
\,.
\tag 3.23
$$

We are left with the proof of the Peierls condition
(3.16). We start with the observation that for
$(U,\mu)\in S_+^{(\epsilon)}\subset S_+$
and arbitrary values for the spins
$S_\bi$ and $S_\bj$,
$$
h(S_\bi,S_\bj)
\geq
\min\{h(0,+),h(-,+)\}
\,,
\tag 3.24
$$
while
$$
h(S_\bi,S_\bj)
\geq
\min\{h(0,+),h(-,+)\}+\alpha\epsilon
\,,
\tag 3.25
$$
for some dimension dependent constant
$\alpha>0$ whenever the bond
$\langle\bi,\bj\rangle$ is excited.
Combining  (3.24) and (3.25)
with the fact that
$$
h_0^{m}(\langle\bi,\bj\rangle)
\leq\min\{h(0,+),h(-,+)\}
\,,
\tag 3.26
$$
we obtain the bound
$$
|z(Y)|
\leq
\sum_{S_Y:\sigma(S_i)=\sigma_i}
\;\prod_{\langle\bi,\bj\rangle\in B^*(Y)}
e^{-\beta \alpha\epsilon}
\,,
\tag 3.27
$$
where $B^*(Y)$ is the set of excited bonds in $B(Y)$.
Bounding $|B^*(Y)|$ from below by
$\half|\supp Y|$, and
the number of terms in the sum
over $S_Y$ from above by $2^{|\supp Y|}$,
we obtain the
bound (3.16) with
$$
e^{-\tau}=2e^{-\half\beta \alpha\epsilon}
\,.
\tag 3.28
$$

Given (3.14), (3.15) and (3.16), the
partition function (3.14) can be expressed
as the partition function of a contour
system with exponentially decaying weights,
$$
Z_m(\Lambda)=
e^{-\beta H_\Lambda^\eff(g^m_\Lambda\mid m)}
\sum_{\{Y_1,\cdots,Y_n\}}
\;\prod_{k=1}^n z(Y_k)
\tag 3.29
$$
where the sum runs over sets
$\{Y_1,\cdots,Y_n\}$ of matching contours
obeying the compatibility condition
that $\dist(\supp Y_k,\supp Y_l) > 1$ for
$k\neq l$.
As a consequence, the model can be again analyzed
by standard methods, see e\.g\. \cite{Zah84}.
One
obtains
the existence of the limits
$$
\langle - \rangle_m
=\lim_{L\to\infty}
\langle - \rangle_{m,\Lambda}
\tag 3.30
$$
as periodic Gibbs states with
$\Delta>0$ and $\Delta<0$,
respectively, the fact that these states
are extremal, and the fact that all
other periodic Gibbs states are convex combinations
of $\langle - \rangle_\even$
and $\langle - \rangle_\odd$.

Considering finally a suitable complex
neighbourhood of the region $S_+^{(\epsilon)}$,
e\.g\.
$$
\Cal S_+^{(\epsilon,\delta)}
=\{(U,\mu)\mid
(\text{Re}\,U,\text{Re}\,\mu)\in S_+^{(\epsilon)}
\quad\text{and}\quad
|\text{Im}\,U|<\delta,|\text{Im}\,\mu|<\delta
\}
\tag 3.31
$$
with $\delta$ sufficiently small,
one easily establishes a Peierls condition
with a slightly smaller Peierls constant
$\tau=\tau(\beta,\epsilon,\delta,d)$.
The methods of \cite{BI89} then give%
\footnote{As in the proof of Proposition 2,
we use the fact that,
by translation invariance,
the corresponding truncated free
energies $f_\even$ and $f_\odd$ are equal in the
whole complex region $\Cal S_+^{(\epsilon,\delta)}$.}
the free energy density $f$
and the staggered charge-order parameters
$\Delta_\even(\beta,U,\mu)$ and $\Delta_\odd(\beta,U,\mu)$
as convergent sums
of analytic terms
in $\Cal S_+^{(\epsilon,\delta)}$,
implying their analyticity in
$\Cal S_+^{(\epsilon,\delta)}$
and hence their real analyticity
in $S_+^{(\epsilon)}$.
This completes the proof of Proposition 4.
\qed
\enddemo

\remark{Remark}
It is intriguing to relate the first order
jump of the staggered order parameter at
$T=0$ to the analytic behaviour at positive
temperatures.
To this end, we note that
the distribution of the spin variable
$S_\bi$  in the
restricted ensembles $\Cal G_m^+$
is given by
$$
\mu(S_\bi)=\cases
       \delta(S_\bi,+)&\text{if}\qquad g_\bi^m=+
       \\
       {e^{-2d\beta h(S_\bi,+)}\over\lambda}
       (\delta(S_\bi,0)+\delta(S_\bi,-))
                      &\text{if}\qquad g_\bi^m=-
\endcases
\,.
$$
The finite temperature excitation above the corresponding
ground states $\sigma=g^m$
slightly modify these distributions, leading to
corrections of the order $O(e^{-\beta\alpha\epsilon})$.
As a consequence, the staggered
order parameter $\Delta$ in the region
$S_+^{(\epsilon)}$ is given by
$$
\align
\Delta
&=
\pm\biggl(
\frac 34 +\frac 14\tanh\Bigl(2d\beta(h(0,+)-h(-,+))\Bigr)
\biggr)
+O(e^{-\beta\alpha\epsilon})
\\
&=
\pm\biggl(
\frac 34 + \frac 14 \tanh\bigl(\beta(2d-\mu-\tilde U/2)\bigr)
\Bigr)
+
O(e^{-\beta\alpha\epsilon})
\,,
\tag 3.32
\endalign
$$
where the plus sign corresponds to the even phase and
the minus sign corresponds to the odd phase.
\endremark

\subhead
Acknowledgment
\endsubhead
  One of the authors (J.J.) would like to express his thanks to
 the organizers of the miniworkshop, held at
 the Center for Theoretical Study
 of the Charles University in Prague, for  invitation.
The present paper arose from discussions held at this meeting.

\vfil\eject

\enddocument
\end